# First-principles Quantum Insights into Bandgap Engineering, Valley Quantum Hall Effect, and Nonlinear Optical Response of Ge-Doped Graphene for Potential Optoelectronic Applications

Sana Maroof [a†], Abdul Sattar [a*†], Azmat Iqbal Bashir [b], Muhammad Irfan [c], Hamid Latif [d], Hina Mustafa[a], Ahmad Saeed [a], Raja Junaid Amjad [a], Farah Alvi [a]

[a]Physics Department, COMSATS University Islamabad, Lahore Campus, Defence Road, Lahore 54000, Pakistan
[b]Department of Physics Faculty of Engineering and Applied Science, Riphah International University, Islamabad, Pakistan
[c]Department of Engineering Physics, Shenzhen Technology University, Shenzhen, 518118, China
[d]Department of Physics, Forman Christian College University Lahore 54000, Pakistan
[†]Co-first authors, Authors have equal contribution
[*]Corresponding author: asattar@cuilahore.edu.pk

## Abstract

The valley in the band structure of materials has gained a lot of attention recently. The promising applications of the valley degree of freedom include the next-generation valleytronic devices, quantum information processing, quantum computing, and optoelectronic devices. Graphene is an ideal quantum material for high-speed valleytronic applications because of its high carrier mobility and convenience of bandgap engineering. Employing first-principles density functional theoretical approach, this study opted bandgap engineering strategy via Germanium doping to open bandgap and enhance valley selectivity in graphene monolayers. The impact of Ge dopant concentration of 2%, 3.125%, 5.5%, and 12.5% is explored on the valleytronic; valley Hall effect, valley transport, and optical properties. The reported results demonstrate that bandgap, valley polarization, and second harmonic generation can be tuned effectively by varying doping concentration of Germanium. The Berry curvature profile is antisymmetric for corresponding K and K′ valleys, thus leading to valley-dependent transport properties and a potential valley Hall effect. Finally, the second-order susceptibilities exhibit corresponding optical absorption peaks, indicating efficient second-harmonic generation due to the broken inversion symmetry. These findings highlight the potential of Ge-doped graphene for nonlinear optics and valleytronics applications, while providing novel insights into its topological phase and transport properties.

## 1. Introduction

The quantum degrees of freedom (QDOF) refer to various quantum variables such as electric charge, orbital and spin angular momentum, spatial parity (inversion of spatial coordinate), and chirality (the lack of invariance of an object under reflection), helicity, and so on. These QDOF have played a significant role in the understanding of working principles and development of conventional electronics, optical, and optoelectronic devices and modern quantum technology [1,2]. The properties and functionality of quantum materials can be tuned effectively by manipulating the QDOF. For instance, electric charge, orbital and spin angular momenta, play an essential role in understanding the quantum origin of quantum Hall effect, superconductivity, and magnetic properties of materials. Strategy of spin-orbit coupling is adopted to understand and tune the magnetic aspects of materials. Nature and topology of energy bands can be also considered as potential tunable QDOF by chemical doping or by applying external fields, which is the hallmark of novel quantum materials such as topological insulators and Weyl (semi)metals [3]. In condensed matter physics, electronic band structure of crystalline materials is the relation between energy of an electron and its momentum in crystal. Generally, a valley is referred to as a local maxima and minima in the valence band and conduction band, respectively. The electronic properties of ideal 2D valley materials, with honeycomb hexagonal crystalline structure such as graphene, are dominated by the two spatial symmetric valleys (maxima and minima) or edge states overlapping at the Fermi level, at the band edges in the Brillouin zone, called the Dirac K-points ($\pm K$). However, when this spatial inversion symmetry is broken while keeping the time-reversal symmetry in intact by external perturbation such as fields or doping, a bandgap is created in otherwise gapless materials such as graphene. This results in the creation of indistinguishable energy-degenerate valleys at different K points along the crystal axes. Besides, a valley-contrasting opposite curvatures in the energy bands are induced, called the Berry curvatures. The curvatures act like effective magnetic fields in the momentum space and cause valley-contrasting Hall effect. This causes a flow of charges and hence the current towards the edges of the material. However, under the normal situation of spatial inversion symmetry of electron valleys in the gapless materials, there is no such flow of carries and hence no Hall current.

These maxima/minima in the energy bands, called valley degree of freedom (VDOF), has motivated immense research interest. Analogous to the role of spin in spintronics, valleytronics is the branch of electronics that exploits the possible potential of the VDOF of electrons for fundamental prospects and technological implications. The VDOF of electrons/holes can be tuned effectively and exploited to process, store, and encode information for promising applications in the emerging fields of quantum optics, nanophotonics, quantum information, quantum communication, and quantum computing. In recent decades, the concept of valleytronics [4] employing the honeycomb geometrical lattices, such as graphene and WSe2, has motivated an immense theoretical and experimental research attention for various fundamental and practical applications [5–12]. The electrons valleys in the energy bands can also be interpreted as a binary pseudospin (carriers) pair analogous to a spin-1/2 system such that the electrons in the $\pm K$ valleys can be categorized as valley-pseudospin up and valley-pseudospin down. Analogous to spin Hall effect, this phenomenon of intriguing interest is also called the valley Hall effect as it causes the electrons/holes of the opposite valleys to flow to the opposite edges of materials. The divergent character of carriers, and hence the associated current, to move away from each other resembles magnetic flux, called the Berry curvature.

For electrons in crystals, the momentum-dependent periodic Bloch wave function $|u(k, r)\rangle$ in a periodic crystal defined by $|\Psi(k, r)\rangle = \exp[ik \cdot r]|u(k, r)\rangle$, satisfies the time-independent eigenvalue Schrodinger equation in momentum space $H(k)|u(k)\rangle = \langle u(k)|u(k)\rangle$. To understand the energy-degenerate electron states of the valleys, and the induction of the Berry curvature, consider an overlapping of an eigenvector at point k and a nearby point k + δk in momentum space, then by Taylor expansion

$$\langle u(k)|u(k + \delta k)\rangle \approx 1 + \delta k \cdot \langle u(k)|\nabla_k|u(k)\rangle \approx e^{iA(k) \cdot \delta k} \quad \ldots\ldots\ldots\ldots\ldots\ldots\ldots\ldots(1)$$

where $A(k) = -i\langle u(k)|\nabla_k|u(k)\rangle$ is called the Berry connection and acts as a local gauge potential whose curl $\Omega(k) = \nabla_k \times A(k)$ is called the Berry curvature, which is analogous to magnetic field defined as $B = \nabla \times A$, where $A(k) \cdot \delta k$ has is the corresponding geometrical phase [11]. The flux of the Berry curvature over any closed surface (e.g. 2D Fermi surface) is quantized as given by the Chern theorem $1/2\pi \int \Omega(k)(k) \cdot dS = \pm C$, where the integer C is called Chern number.

According to Gauss' theorem, for a nonzero flux of the Berry curvature through a Fermi surface sheet, there must be a point sink, or a point source associated with the Berry curvature. This point sink or source of the Berry curvature is a clear indication of a singular point or a Valley in the electronic band structure. This singularity in turn represents a degeneracy between two energy bands, called a Weyl node or nodal point. Whereas the Berry curvature in the vicinity of the Weyl nodal point corresponds to a monopole. The energy band Hamiltonian for a spatially symmetric systemic expanded in the vicinity of the Weyl node is expanded as

$$H(k) = \pm v_F \sigma \cdot k \quad \ldots \ldots \ldots \ldots \ldots \ldots \ldots \ldots \ldots \ldots (2)$$

where $v_F$ is the Fermi velocity at the Dirac point, σ represents Pauli matrices, the signs ± represent the topological charge C of the Weyl node. This Hamiltonian corresponds to the massless relativistic fermions for a gapless graphene. The topological charge represents the chirality of Weyl fermions, for a right-handed (R) chiral fermion C = +1 and for a left-handed (L) chiral fermion C = −1.

Now for a system undergoing spatial-inversion symmetry breaking, owing to spin-orbit-coupling or otherwise (e.g. doping or spontaneous magnetization) the mass term is generated, and the modified Hamiltonian is reformulated as follows [13] (and references therein),

$$H(k) = \pm \hbar v_F \sigma_i \cdot k_i + \hbar m \sigma_z \quad \ldots \ldots \ldots \ldots \ldots \ldots \ldots \ldots \ldots (3)$$

For a 2D graphene,

$$H(k) = \pm \hbar v_F (\sigma_x k_x + \sigma_y k_y) + \hbar m \sigma_z \quad \ldots \ldots \ldots \ldots \ldots (4)$$

where $\sigma_z$ is the z-component of electron spin

The phenomenon is also known as anomalous quantum Hall effect.

In the case of the valley quantum Hall effect, the Hamiltonian of 2D graphene [14] is defined by

$$H(k) = \hbar v_F (l \sigma_x k_x + \sigma_y k_y) + \hbar l k_x \sigma_0 + \hbar m \sigma_z \quad \ldots \ldots \ldots \ldots \ldots \ldots \ldots (5)$$

Where $l = \pm$ is for the valley index.

The developments in valleytronics have paved the way for the realization of novel applications of 2D (two-dimensional) materials that include valley-based transistors, photodetectors, information storage and quantum computing [15-17]. Valleytronics exploits VDOF to tune the functionality and properties of 2D materials for potential basic and applied implications. Among the 2D novel materials, graphene has been a center of attention of the scientific community for more than two decades due to its unmatchable characteristics such as high carrier mobility, excellent mechanical strength, tunable non-linear optical properties, and ease of fabrication and integration in current semiconductor devices. However, pristine graphene has zero bandgap which is one of the biggest hurdles to use in optical, electronic or optoelectronic devices. To overcome this problem, various bandgap engineering strategies have been applied such as doping or hybrid structure of graphene with similar materials [18,19]. Moreover, pristine graphene does not exhibit intrinsic valleytronic properties because its inversion symmetry preserves the degeneracy of the K and K' valleys. In the past, these problems have been tackled using various ways including doping, strain application, introduction of heterolayers and molecular functionalization [12,18-22]. There are a variety of dopants available for graphene such as B, N, S, P, Ga, and Ge [23]. Among these dopants, Ge has been proven to be useful for opening bandgap [24]. The introduction of spin-orbital-coupling (SOC) also breaks the intrinsic inversion symmetry of graphene. Recent research has highlighted the potential of doped graphene with elements like Gallium, Germanium, Arsenic, and Selenium to open a bandgap ranging from 0.3 to 1.3 eV [24], which is significant for valleytronics applications. Although bandgap opening due to Ge doping has been studied in the past, systematic investigation on various effects, such as SOC, Berry phase, valley selective transport, and nonlinear responses, on bandgap engineering are not fully explored in this quantum system.

This study aims to fill the existence research gap by reporting a systematic investigation on the role of Ge doping concentration on the bandgap engineering of graphene, tuning of the Berry curvature profiles, valley Hall effect, and second harmonics generation. For the first-principles density functional theory-based calculations, Generalized Gradient Approximation (GGA) within the Perdew-Burke-Ernzerhof (PBE) parameterization and the hybrid functional HSE06 (Norm Conserving Scalar relativistic pseudopotential) were employed with and without SOC coupling. This approach of Ge doping into graphene with application of SOC coupling will enhance our understanding on the fundamental aspects of the VDOF considering the Berry curvature and valley Hall effect for potential applications of valletronics in the emerging fields of quantum information and communication, quantum computing, and advanced optoelectronic devices.

The layout of the study is presented in the following order. Section 2 is focused on the key theoretical formalism based on density functional theory. Interpretation and analysis of the obtained theoretical results is accomplished in Section 3. The conclusion of the study with key findings and future prospects is presented in Section 4.

## 2 Materials and computational methods

For the computation of the target properties, first-principles Density Functional Theory (DFT) calculations were performed using the plane-wave pseudopotential approach, as implemented in the open-source Quantum ESPRESSO package [25]. Electronic occupations were treated using the cold smearing method with a smearing width (degauss) of 0.01 eV.

Stringent convergence criteria were adopted to ensure accurate results. The total energy convergence threshold was set to $2.0 \times 10^{-5}$ eV, the force convergence threshold to $1.0 \times 10^{-4}$ eV/Å, and the electronic self-consistency convergence threshold to $4.0 \times 10^{-10}$ eV. The charge density mixing parameter was fixed at 0.4, and the maximum number of electronic iterations was limited to 100.

Structural optimization was carried out using the Plane-Wave Self-Consistent Field module (pw.x), with relaxation of atomic positions, lattice parameters, plane-wave cutoff energies, and k-point sampling. The kinetic energy cut-off for the wavefunctions was set to 40 Ry, while a cutoff of 400 Ry was used for the charge density. A vacuum spacing of 20 Å was applied along the z-direction to eliminate interactions between periodic images in the two-dimensional (2D) system.

Brillouin zone sampling was performed using a Monkhorst-Pack k-point mesh of 12×12×1 for pristine and doped graphene supercells. The exchange-correlation energy was treated within the generalized gradient approximation (GGA) using the Perdew-Burke-Ernzerhof (PBE) functional. To improve accuracy, hybrid functional calculations using HSE06 were also conducted with norm-conserving pseudopotentials. SOC effects were included via fully relativistic pseudopotentials to capture valley-dependent and topological phenomena.

Electronic properties were computed through self-consistent field (SCF) calculations, followed by band structure analysis using Band Structure Post-Processing (band.x), and density of states (DOS) computations using dos.x. Optical properties, including second harmonic generation (SHG), were evaluated using ph.x and relevant post-processing tools. For SHG, SCF and non-self-consistent field (NSCF) calculations were performed, followed by Wannierization using Wannier90, and SHG coefficients were extracted using EPW and custom scripts. Topological characteristics, such as Berry curvature and the Chern number, were investigated using the Quantum ESPRESSO-Wannier90 interface. The Berry curvature was calculated in the Brillouin zone using postw90, and the resulting data were analysed with Python scripts to determine the Chern number.

*Graphene and Germanium Doping*

Graphene, a monolayer of carbon atoms arranged in a two-dimensional honeycomb lattice, exhibits a hexagonal Brillouin zone characterized by two inequivalent Dirac points, K and K′. These points correspond to reciprocal momenta of massless Dirac fermions and form the vertices of conical energy-momentum dispersion structures. The energy bands intersect at the Fermi level, resulting in a linear dispersion near the Dirac points.

To modulate the electronic and topological properties of graphene, substitutional doping with germanium was introduced. Doping concentrations of 12.5% Ge, 5.56% Ge, 3.12% Ge, and a lower unspecified concentration were modelled by replacing a single carbon atom with a germanium atom in 2×2×1, 3×3×1, 4×4×1, and 5×5×1 supercells, respectively (Figure 1). These doped configurations were structurally relaxed and analysed using the same computational workflow as for pristine graphene. The impact of doping on the electronic structure was examined through calculations of the bandgap, density of states, Berry curvature, optical absorption, and SHG response. Both spin-polarized and non-spin-polarized cases were considered, and all structural models were visualized and analysed using XCrySDen.

The results of these simulations are presented and discussed in the following section.

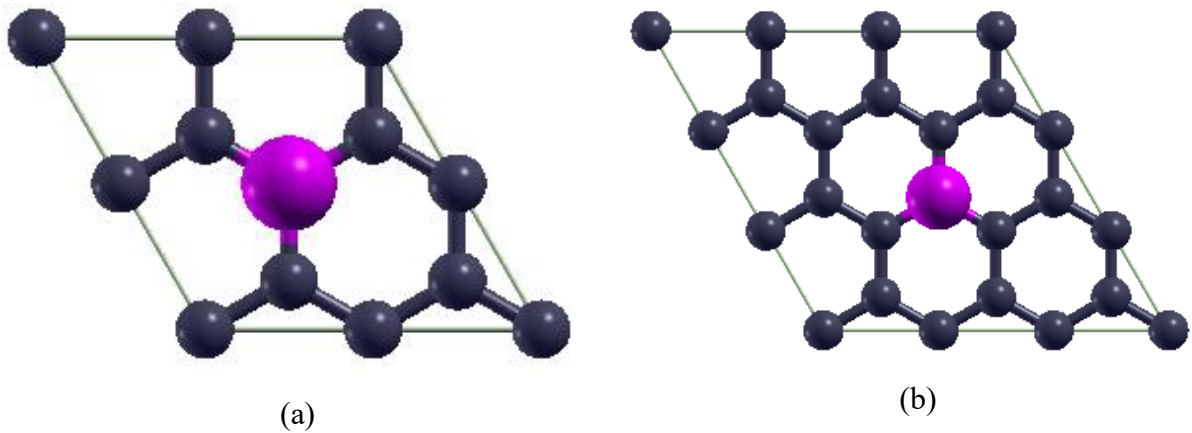

(a)  (b)

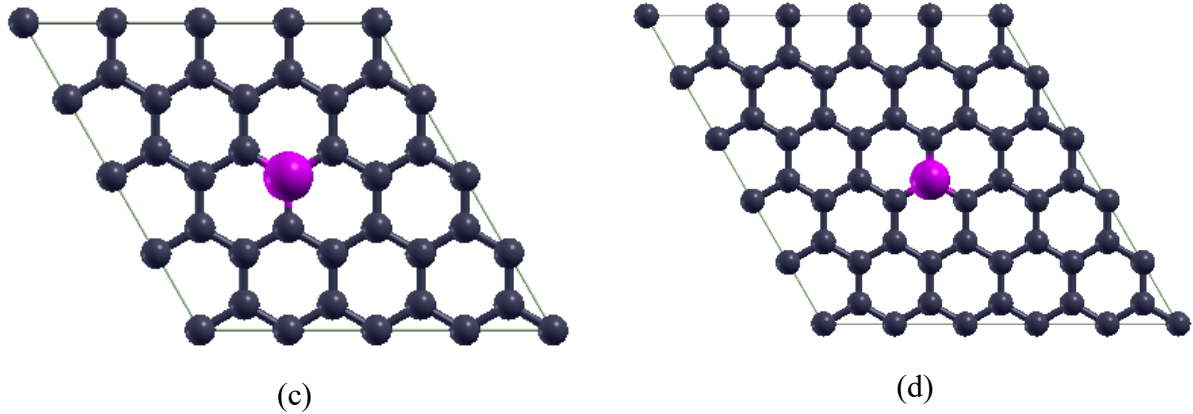

(c)                                         (d)

Figure 1: Germanium-doped supercells of monolayer graphene at different concentrations. (a) graphene with 12.5% Ge doping in a 2×2×1 supercell. (b) graphene with 5.56% Ge doping in a 3×3×1 supercell. (c) graphene with 3.12% Ge doping in a 4×4×1 supercell. (d) graphene with 2% Ge doping in a 5×5×1 supercell. Grey spheres correspond to carbon atoms, and purple-red spheres denote germanium atoms.

To set the bench mark, band structure of the pristine graphene was calclated as shown in Figure 2. The Dirac points are visible at K and K′ points, and the computed bandgap of the prinstine graphene is zero.

In this investigation, the band structure of pure graphene was calculated along the Γ-K-M-Γ trajectory of the momentum and Γ-K-M-K-Γ trajectory within the Brillouin zone. For the pristine monolayer graphene, the conduction band minimum (CBM) and the valence band maximum (VBM) are located at the K-points (Figure 2a) with inversion symmetry intact, indicating that the pristine monolayer graphene is a zero-bandgap semiconductor and hence is a semimetal. In Figure 2a, the valence and conduction bands overlap at the Dirac point (K, K′) having almost identical valleys point. Figure 2b shows the overlapping of valence and conduction bands at two different values of momentum forming the Dirac points but the spatial inversion of bandgap symmetry is again preserved, which otherwise needs to be broken for the induction of the energy-degenerate valley states.

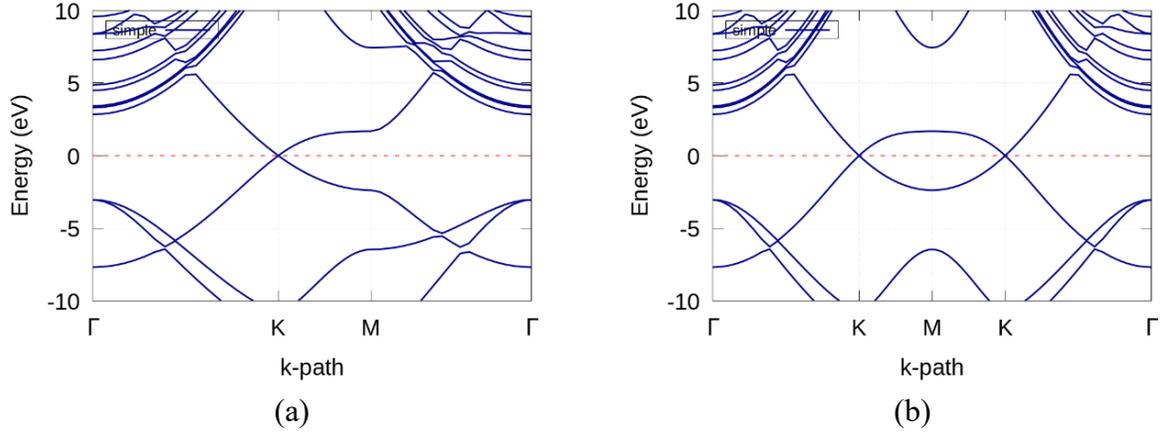

(a)                                  (b)

Figure 2: Band structure of the pristine graphene calculated along high-symmetry k-points in the hexagonal Brillouin zone: (a) Γ-K-M-Γ and (b) Γ-K-M-K-Γ, illustrating the inversion symmetric linear dispersion near the Dirac points at K and K'.

Figure 3: elucidates the electronic band structure of graphene doped with one atom of Germanium. The calculations were performed using both the Generalized Gradient Approximation (GGA-PBE) and the hybrid functional HSE06 (Norm Conserving Scalar relativistic pseudopotential) without considering spin effects.

The introduction of Germanium into the monolayer of graphene leads to the opening of bandgap as shown in Figure 3 as well as the breaking of inversion symmetry of the Dirac valleys. However, the energy-degenerate time-reversal symmetry of the two prominent valleys is preserved that will be discussed later in this paper. In particular, two pairs of contrasting valleys are produced such that for each couplet one valley lies above and the other below the Fermi level. It is observed that the width of bandgap increases with the increasing concentration of Ge doping. For instance, at a higher doping concentration of 12.5%, the Hall valleys become more prominent and appear at the K and K′ points (Figure 3a). Interestingly, as the doping concentration increases to 5. 5%, the band structure does not show distinct valleys. Although, the bandgap along the orientation Γ-Γ is opened considerably, the Dirac nodal points shift to Γ-points (Figure 3b). With further decrease in doping concentration to order 3.125% (Figure 3c) and 2.0% (Figure 3d), the bandgap topology of the samples is modified considerably. However, the bandgap of the valley doublet is decreased at the points (K, K′).

Our next strategy is to investigate the impact of SOC coupling on the bandgap opening and the valley Hall effect. SOC coupling is a quantum mechanical effect that measures the interaction strength between the orbital and spin motion of electrons in the crystal. This is usually expressed as the scalar product of L and S: $L.S = 1/2(J^2 - L^2 - S^2)$, where $J$ is the total angular momentum. This effect is particularly significant in materials containing heavy elements, where relativistic influences become more pronounced due to a higher atomic number with unpaired electrons. Graphene, composed of light carbon atoms, exhibits negligible SOC. However, doping with heavier elements like Germanium can introduce or enhance SOC. In this study, we analyzed supercells of sizes Fig.4(a) 2×2×1, Fig.4(b) 3 × 3 × 1, Fig.4(c) 4 × 4 × 1 and Fig.4(d) 5 × 5 × 1, each containing a single substitutional dopant atom. The calculations were carried out using the hybrid functional HSE06 with fully relativistic pseudopotentials and the results for SOC are presented in Figure 5. Incorporating SOC in calculations could provide deeper insights into the interplay between doping, spin polarization, and valley-dependent Hall effect in Germanium-doped graphene. As can be seen, the implication of SOC results in the lifting of the orbital degeneracy of the bands at the K(K′) points. This phenomenon in graphene has been also studied in the past [26].

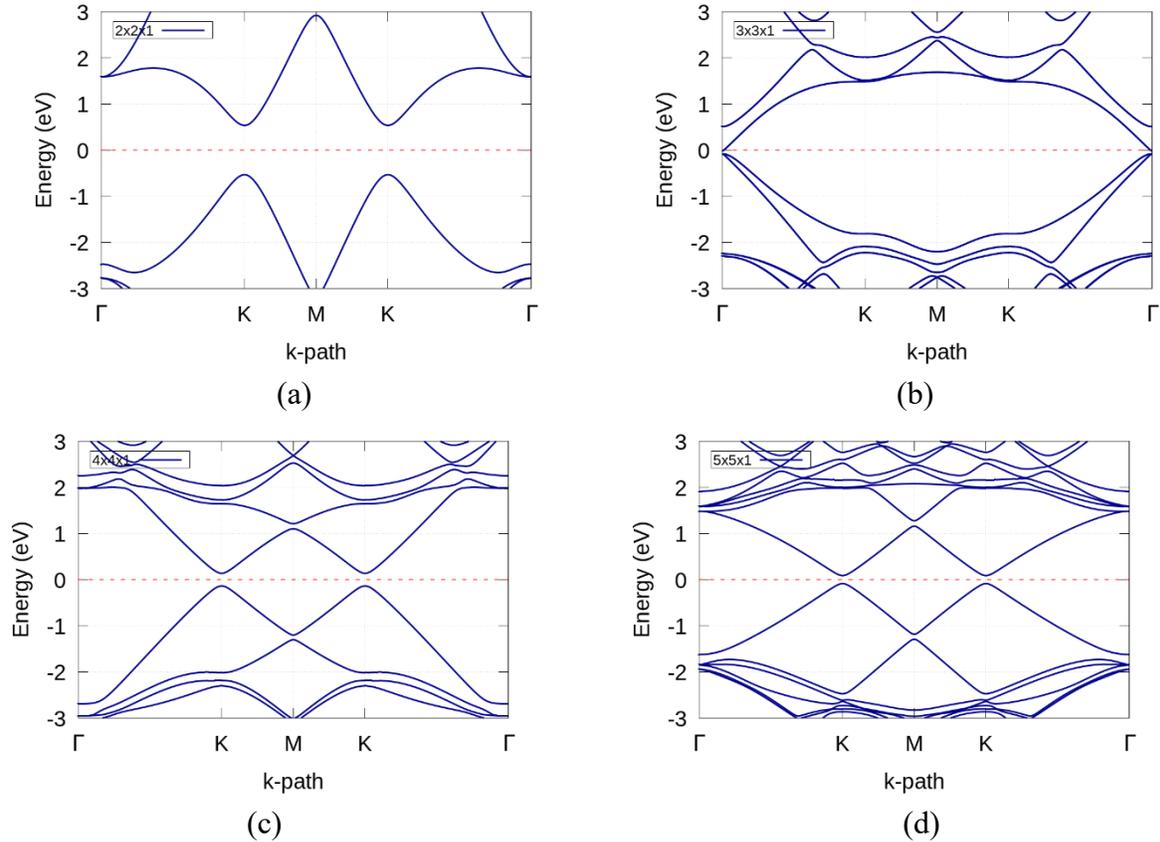

(a)

(b)

(c)

(d)

Figure 3: Band structure of Germanium-doped graphene with doping concentrations of (a) 12.5%, (b) 5.5%, (C) 3.125%, and (d) 2%, calculated using the Generalized Gradient Approximation (GGA) functional without spin-orbit coupling (SOC).

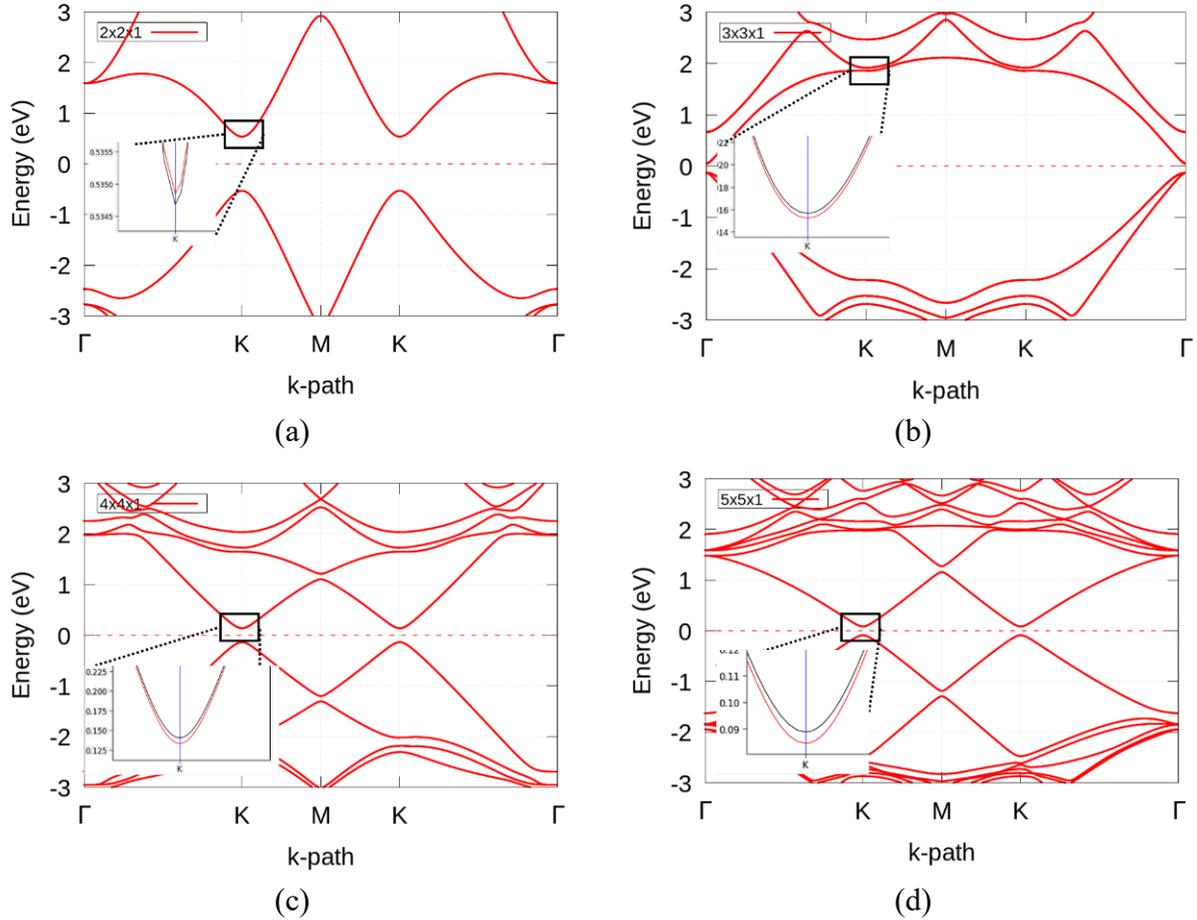

Figure 4: Band structure of Germanium-doped graphene with doping concentrations of (a) 12.5%, (b) 5.5%, (c) 3.125%, and (d) 2%, calculated using the HSE06 functional with spin-orbit coupling (SOC). The black line represents spin-down states, while the red line represents spin up states.

The band structures of pristine and germanium-doped graphene were analyzed along the $\Gamma - K - M - \Gamma$ and $\Gamma - K - M - K' - \Gamma$ trajectories in the Brillouin zone. For pristine graphene, the valence and conduction bands touch at the *K* point, confirming its zero-band-gap nature. Germanium doping opens a band gap, as summarized in Table 1.

Table 1: The calculated bandgaps (eV) of Germanium-doped graphene with different supercell sizes, computed using GGA and Hybrid functionals in polarized and unpolarized states.

| Supercell | GGA Unpolarized | GGA Polarized | Hybrid Unpolarized | Hybrid Polarized | References |
|---|---|---|---|---|---|
| 2 × 2 × 1 | 0.79 | 0.79 | 1.06 | 1.06 | [4] |
| 3 × 3 × 1 | 0.00 | 0.00f | 0.00 | 0.168 | |
| 4 × 4 × 1 | 0.184 | 0.183 | 0.273 | 0.26 | |
| 5 × 5 × 1 | 0.120 | 0.11 | 0.17 | 0.168 | [5] |

The corresponding plots for the calculated density of states were analyzed and presented in Figure 5. The identical DOS profiles for spin up and spin down electrons show that there is no net spin polarization and the magnetic moment of the systems is zero.

However, these results confirm that the doping of Ge in graphene facilitates to open a considerable bandgap. The SOC coupling is too minuscule both in the pristine and doped graphene that the valley polarization of these spin levels in conduction band is hard to utilize for addressing the individual valley by using right or left circularly polarized light [15].

To confirm the valley polarization, we calculated the Berry curvature $\Omega_n$ whose distributions in the hexagonal Brillouin zone for the doped graphene are shown in Figure 6. The 2D colour plot shows clearly that the adjacent valleys polarization. The variation of the Berry curvature in momentum space path Γ-K-M-K′-Γ is presented in Figure 7. The first plot for the Berry Curvature in Figure 7a represents a 2 × 2 × 1 germanium-doped graphene supercell, the second plot in Figure 7b corresponds to a 3 × 3 × 1 supercell, and the third plot in Figure 7c represents a 4 × 4 × 1 supercell, and the last one in Figure 7d corresponds to 5 × 5 × 1 supercell.

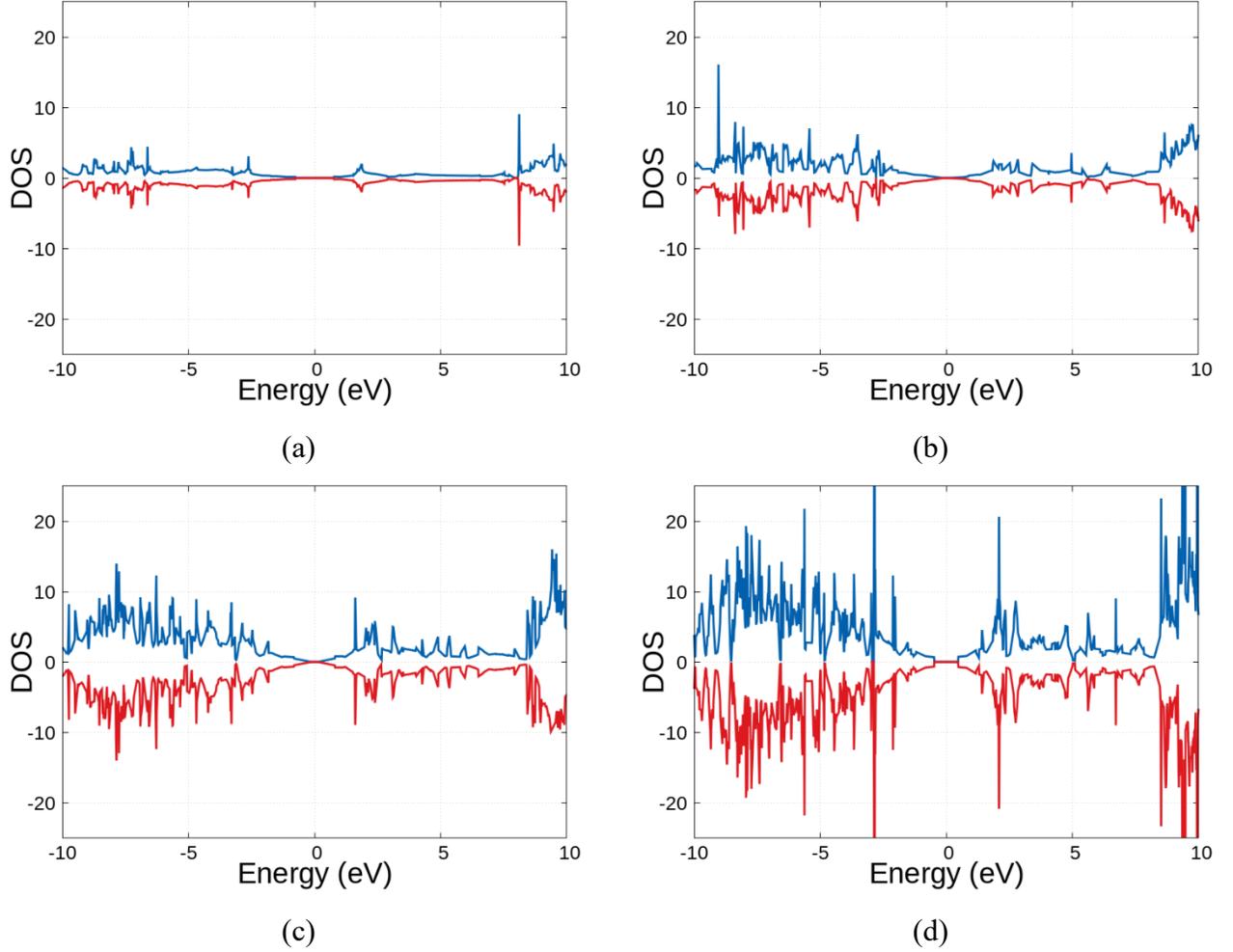

Figure 5: Density of States (DOS) for Germanium-doped graphene with spin polarization at various doping concentrations: (a) 12.5%, (b) 5.5%, (c) 3.125%, and (d) 2%. The spin-resolved DOS illustrates the contributions from spin-up and spin-down states, highlighting the impact of doping concentration on the electronic and spin-dependent properties of the material.

The sizable difference in berry curvature $\Omega_n$ occurs at the $K/K'$ valleys with opposite signs, and the difference in magnitude implies the typical valley contrast characteristic in monolayer graphene. This behavior arises due to breaking of spatial inversion symmetry in momentum space in monolayer graphene due to the doping, i.e., $\Omega_n(k) = \Omega_n(-k)$. The charge carriers acquire an opposite Berry curvature at the $K$ and $K'$ points, as illustrated in the figures. This clearly indicates that the inversion symmetry in the system is broken while the time reversal symmetry i.e. $\Omega_n(k) = -\Omega_n(-k)$ is preserved. On the other hand, in presence of both symmetries $\Omega_n(k) = 0$. Such systems are suitable for valletronic applications [15].

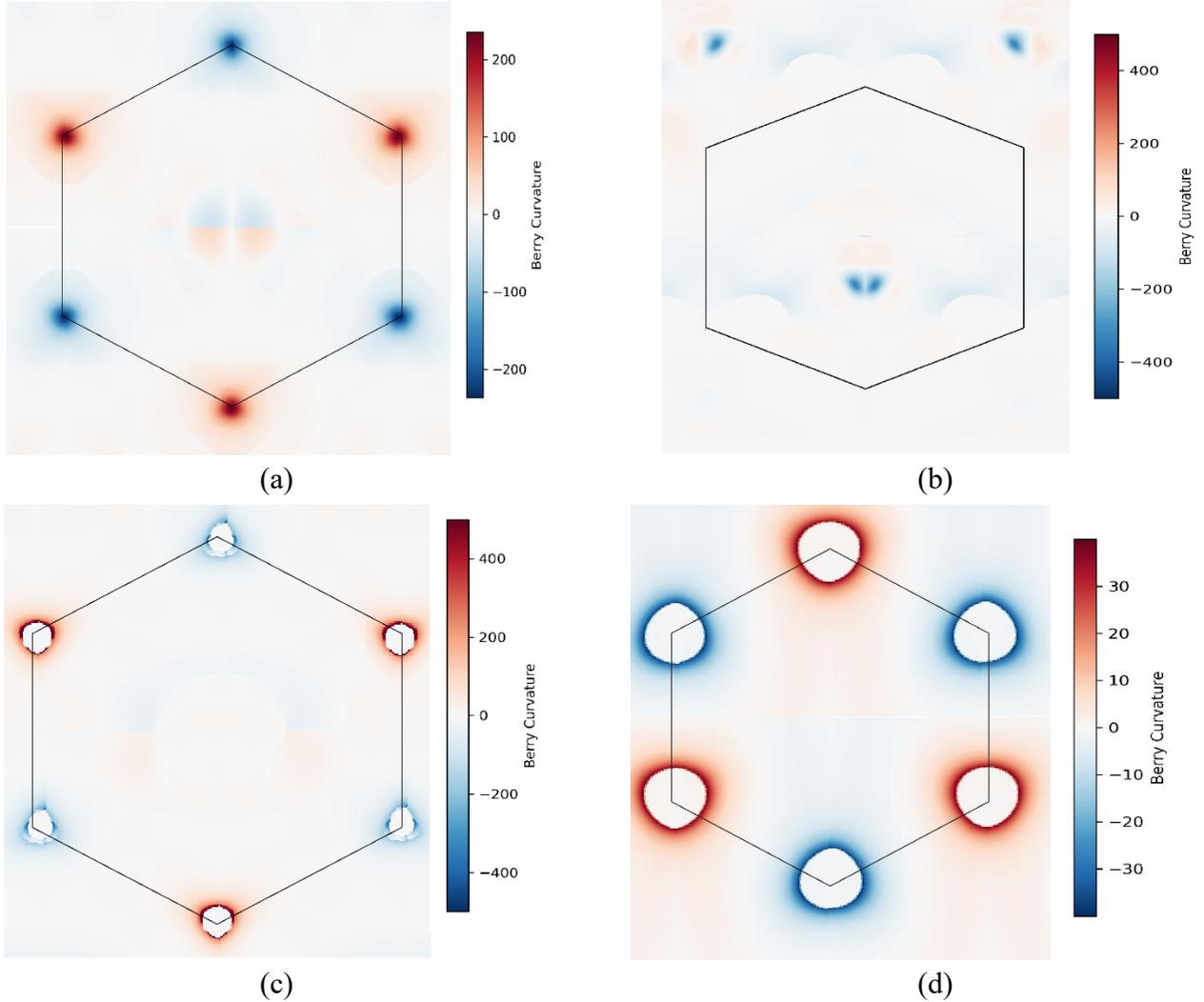

Figure 6: Berry curvature distributions in the hexagonal Brillouin zone for varying Germanium doping concentrations in different supercell configurations: (a) 12.5%, (b) 5.5%, (c) 3.125%, and (d) 2%.

The corresponding Hall velocity is given by $v_h = (e/\hbar)E \times \Omega_n$, which arises from the interband coherence effect. As such, energy bands with larger Berry curvature support fast propagation of carriers. It is understandable because a larger curvature in energy bands leads to smaller effective mass and hence larger velocity. As such the tuning of the Berry curvature can be used to tune the nature of propagation of charge carriers in the materials as depicted by the sketch in Figure 8.

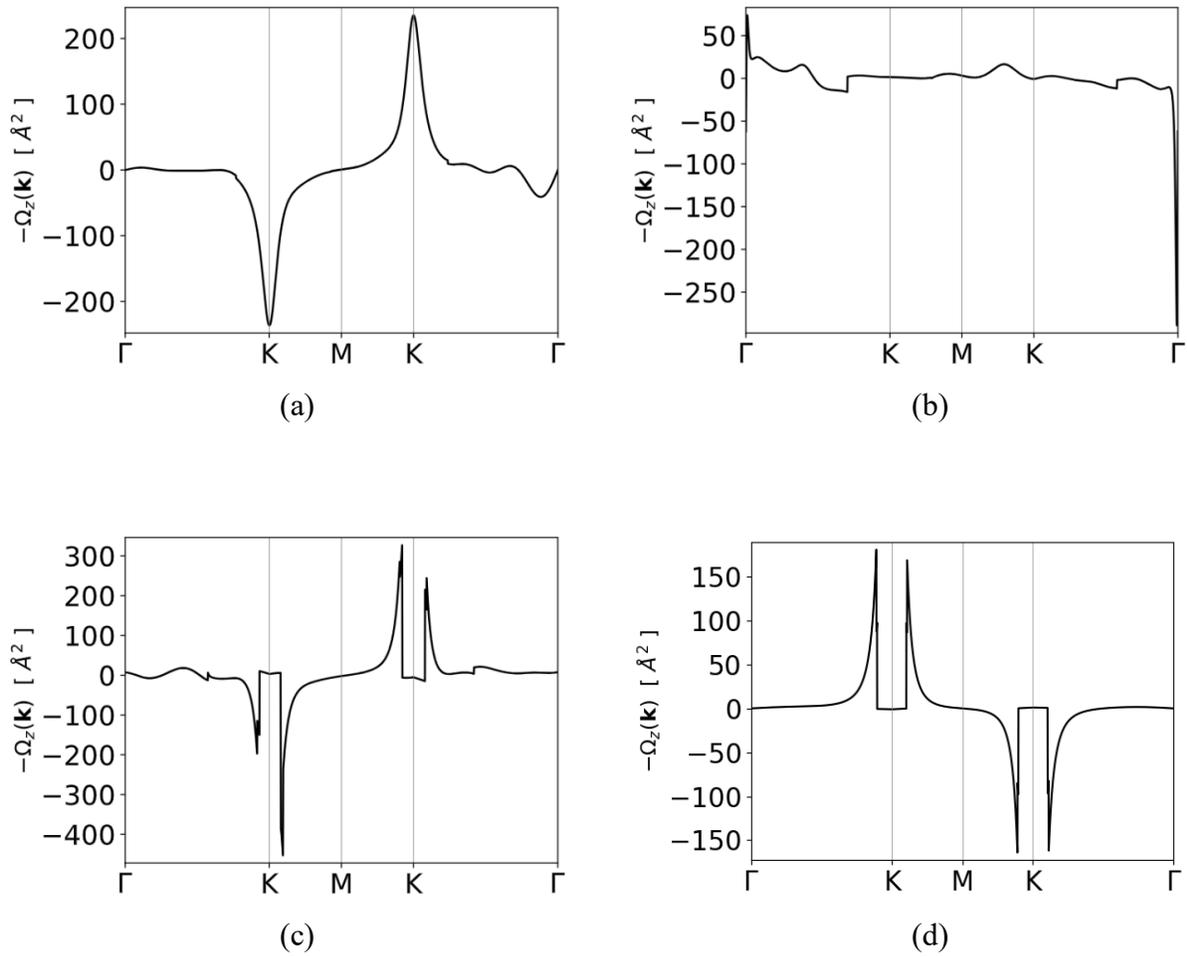

Figure 7: The Berry curvature distributions for varying Germanium doping concentrations in different supercell configurations of graphene: (a) 12.5%, (b) 5.5%, (c) 3.125%, and (d) 2%. The results show valley polarization at the K and K' (K-prime) points, except for the 5.5% doped graphene.

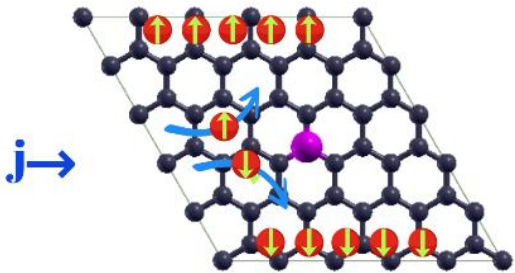

Figure 8: Sketch showing anomalous hall effect produced due to the berry curvature inversion at K and K′ valleys.

The $\Omega_z(k)$ at the $K$ and $K'$ valleys exhibit different magnitudes with opposite signs, suggesting that the valley-contrasting characteristics in germanium-doped monolayer graphene at 12.5% and 3.125% doping remain intact. This also clearly indicates the presence of anomalous Hall conductivity in all systems except 3×3×1. Also, the integration of $\Omega_z(k)$ over the 2D Brillouin zone provides us with Chern number for these systems as presented in Table 2. All systems are trivial topological materials as the Chern number is close to zero for all doping.

Table 2: Chern number values for different supercell sizes and Dopant percentages

| Supercell Size | Ge (%) | Chern Number (C) |
|---|---|---|
| 2×2×1 | 12.5 | 0.0000833 |
| 3×3×1 | 5.5 | -0.0057667 |
| 4×4×1 | 3.125 | -0.000435 |
| 5×5×1 | 2.0 | -0.0007353 |

Another evidence of the inversion symmetry breaking, valley Hall effect, and the Berry curvature, is the observation of the second harmonic generations (SGH) in these dopped monolayers of graphene. In this context, the results on the second order nonlinear susceptibilities $\chi^2$ are obtained. Second harmonic spectroscopy is also an important strategy to directly investigate the valley polarization through emission of second harmonic (SH) light even in quantum materials, such as graphene, processing the centrosymmetric crystals. This can be accomplished by calculating the nonlinear response and the respective SH as a function of electron density, polarization, and degree of valley polarization [27].

SHG is a coherent optical process of radiation of dipoles which oscillate with the external electric field of frequency $\omega$ by radiating electric field of frequency $2\omega$ as well as $\omega$. It corresponds to the second order expansion of polarization, $P(\omega) = \chi^2 E^2$. Because of the inversion symmetry, materials with centrosymmetry do not show SHG. Therefore, the presence of SHG in our proposed system clearly demonstrates the breaking of inversion symmetry.

The computed results for the doped samples of graphene show prominent peaks in $\chi^2$, thus indicating inversion symmetry breaking that might also be due to valley polarization. However, the graphene with 2% doping shows a distinct peak between 0.7 eV to around 1.3eV (see Figure 9). The valley states having less energy bandgap may facilitate the correspondence absorption peaks while lifting electrons from the maxima of valence band to the corresponding minima of the conduction band which are further elaborated from the calculated absorption spectra (Figure 10).

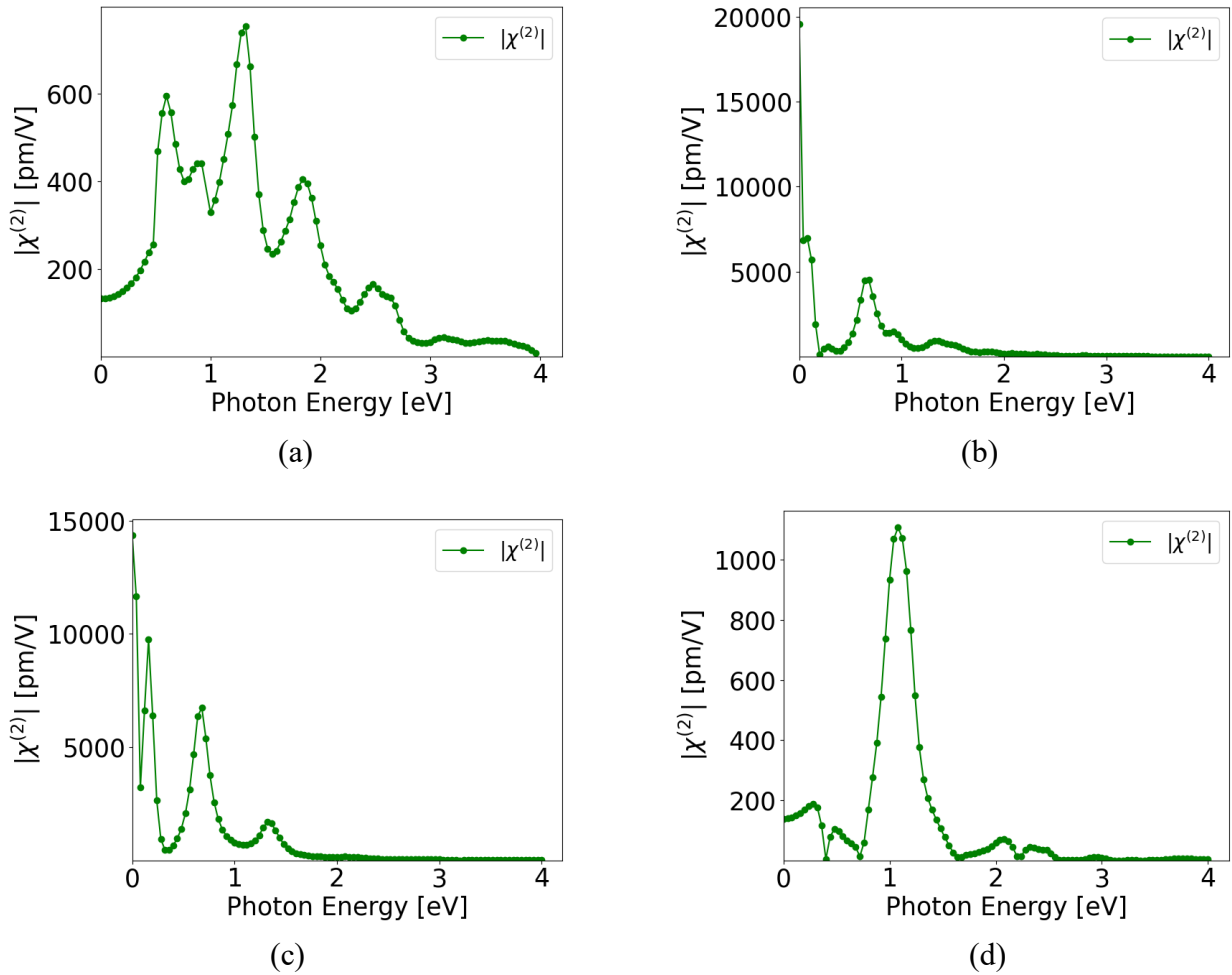

Figure 9: Second Harmonic Generation (SHG) results showing variation in $\chi^2$ with photon energy for (a) 12.5%, (b) 5.5%, (c) 3.125%, and (d) 2% doping.

For further clarification and understanding of the underlying optoelectronic phenomenon in discussion, results on optical absorption spectra were also obtained and are presented in Figure 10. As can be observed, the peaks in the $\chi^2$ graph correspond nicely to the peaks visible in the absorption curves plotted in Figure 10. The peaks in the calculated absorption spectra are consistent with the valley states presented in Figure 5. Although all systems show the SHG response yet system with least doping of order 2% (5×5×1) shows a very distinct peak at 1.1 eV.

Circularly polarized light couples to the valleys charge and the valley current (partly) [12] and hence is a robust method to identify and tune the valley Hall effect and to control the optoelectronic properties of valletronic devices. This may lead to promising potential in the field of ultrafast light-matter interaction. Moreover, the same phenomenon may cause the induction of valley currents on time scales that challenge quantum decoherence in matter.

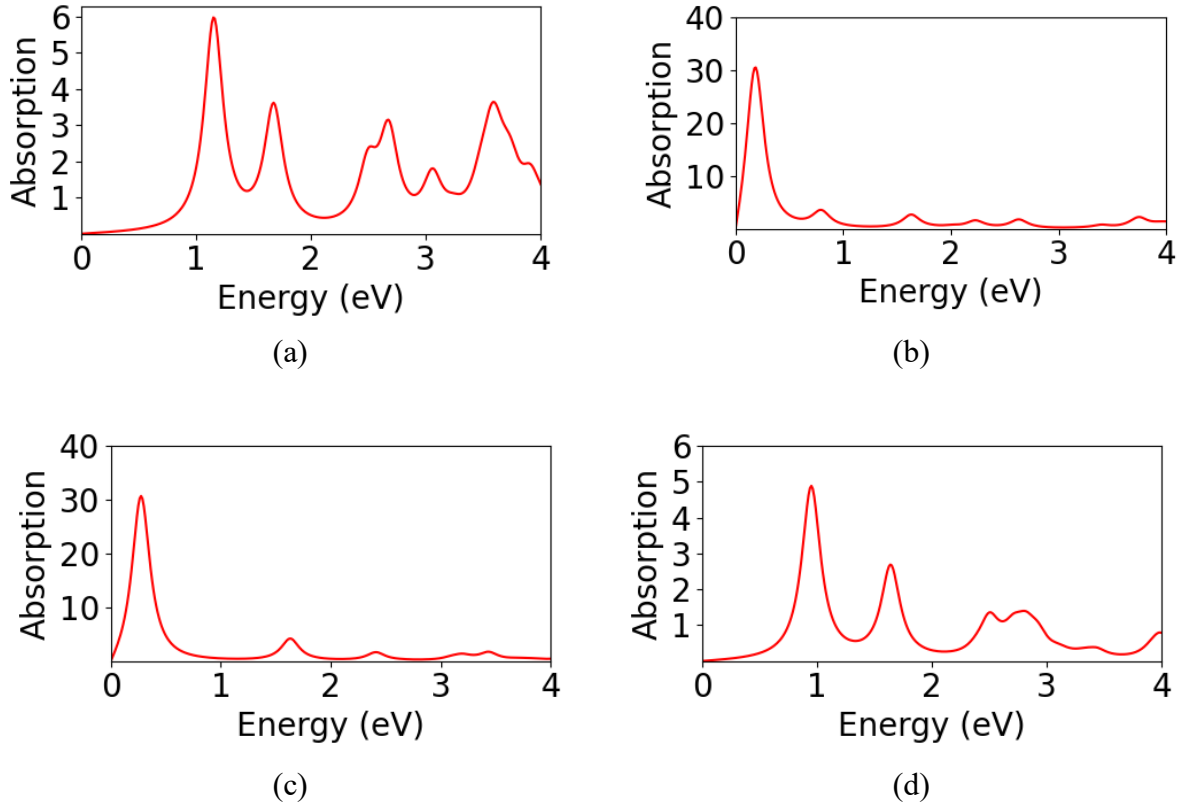

Figure 10: Absorption spectrum of Ge doped graphene with doping of (a) 12.5%, (b) 5.5%, (c) 3.125%, and (d) 2%.

Before concluding the discussion, as a final remark, we believe the key findings of this study not only highlight the novel insights on the fundamental physics for deeper understanding of these phenomena, but it also may pay the way for potential of Ge-doped graphene system as a novel scheme for nonlinear optics and valleytronics applications. The reported non-linear phenomena in this study show the promising potential of the composite structures for fundamental and practical application in nonlinear optics, valleytronics, quantum information process, quantum communication, and quantum computing. For quantum computing, the potential of topological materials has been realized in recent decades owing to stable and protected topological states which are immune to external perturbations and hence are potential candidates to solve the problem of decoherence, which is the biggest hurdle in the practical realization of quantum computers. In this context, we propose the valley Hall states may find potential applications, which are yet to fully realize.

## 4. Conclusions

This study puts forth a new proposal of Ge-dopped composite graphene quantum material as a promising candidate to understand and instigate the phenomenon of the quantum valley Hall effect, valley polarization by computing energy band spectra, the Berry curvature, Chern number, and non-linear optical phenomenon such as absorption spectra and second harmonic generation. The reported results were calculated by employing the first-principle DFT methodology based on GGA-PBE potential and hybrid HSE06 functional with and without spin-orbit coupling. A clear display of bandgap opening, lifting of orbital degeneracy, breaking of spatial inversion symmetry, and the valley Hall effect considering the Berry curvature has been demonstrated from the computed energy band spectra. To confirm further the evidence of the Berry curvature and the valley Hall effect, results on second harmonic generation and optical absorption were also obtained and analyzed. The introduction of SOC with Ge dopant in graphene lifts the orbital degeneracy that opens the opportunity of valley-spin coupling. Opposite signs of the computed Berry phase at K and K′ valleys indicate potential valley-dependent transport. The doping of Ge breaks the inversion symmetry; however, preserving the time reversal symmetry as indicated by zero Chern number showing the system being in topologically trivial phase. Interestingly, the proposed quantum system shows prominent peaks in the computed second order susceptibility and absorption spectra, thus leading to prosing potential of material for nonlinear optical applications. The results imply that the Germanium doping strategy is viable for exploring valleytronics in graphene. The valley spin polarization suggests that the spin valleys can be explored using circularly polarized light. The response on the computed second harmonic generation also confirms the broken inversion symmetry and usefulness of the proposed structure in nonlinear optics. This theoretical strategy is found to be very fruitful to open the bandgap of the pristine graphene and in breaking the spatial inversion symmetry while preserving the time-reversal symmetry which plays a significant role in understanding the above-mentioned physical phenomenon of intriguing fundamental and practical applications in quantum optics, quantum information, quantum computing, and valleytronics.